# Validating Computer Security Methods

Meta-methodology for an Adversarial Science

Antonio Roque
MIT Lincoln Laboratory
Lexington, MA

## ABSTRACT
How can we justify the validity of our computer security methods? This *meta-methodological* question is related to recent explorations on the science of computer security, which have been hindered by computer security's unique properties. We confront this by developing a taxonomy of properties and methods. Interdisciplinary foundations provide a solid grounding for a set of essential concepts, including a decision tree for characterizing adversarial interaction. Several types of invalidation and general ways of addressing them are described for technical methods. An interdisciplinary argument from theory explains the role that meta-methodological validation plays in the adversarial science of computer security.

## 1 INTRODUCTION

If we're developing a new security system, theory or technique, how can we be sure that we are in fact increasing our security? We might develop a metric, but then how do we justify that metric? In general, how do we know if we're using the right *methods*?

In computer security, such *meta-methodological* questions have a long history of being considered [29][49][64][80, pg.v] and are part of the search for a foundational science or theory [19][33][50]. Developing such a science is complicated by several issues [35], such as the fact that we in security have adversaries who might invalidate any solution we propose, and that our technological environment is continually changing.

To address this challenging problem we need definitions rigorous enough to support formal, empirical, and technical methods. We will work in computer security's tradition of using *taxonomies* for concepts and classifications [4][42][44]. This paper is particularly inspired by the approach of "Basic Concepts and Taxonomy of Dependable and Secure Computing," [4] which exemplifies the value of securing a solid foundation for addressing difficult technical problems. However, while that paper describes "attributes" of security such as confidentiality and integrity, it does not describe the *properties* of security that complicate meta-methodology. Likewise, that paper describes "means" such as verification and recovery, it does not describe the *methods* in terms of distinguishing technical methods from other methods that provide context we can use for validation. In fact, we know of no taxonomy that characterizes computer security in the way that we need. So this paper will center on deriving these underdeveloped characteristics as shown in Figure 1; this will help us clarify the difficult questions about validating methods.

In Section 2 we will define a set of properties of cyber security and describe their interrelation. In Section 3 we will define several types of methods in cyber security and describe how they influence each other. We will also see at a general level what types of actions are needed to validate a technical method. Finally in Section 4 we will describe procedures for validating theoretical methods. At every step we will adapt existing concepts from a variety of disciplines so as not to 're-invent the wheel.'

From here on we will use the term *cyber*, which has been used in terms such as *cyberspace* and *cyber security*, to refer to computers, electronic communication systems, and related technical infrastructure systems [15, pg. 60]. Although this term is relatively new, it has the advantage of being technology-neutral. Over the coming decades, our technical environment may move away from traditional computers and networks, while keeping the fundamental attributes of controlling, communicating machines. We seek answers that will remain relevant beyond our immediate technological environment.

## 2 PROPERTIES OF CYBER SECURITY

As with all disciplines, cyber security possesses a set of distinctive properties. Herley and van Oorschot [35, pg.107] warn against using these properties as an excuse to avoid identifying a coherent scientific approach. However, this paper is not making such an excuse; instead, it is using cyber security's unique properties to suggest and constrain meta-methodology. As shown in Figure 1, cyber security is Adversarial, Mechanistic, and Contextual.

### 2.1 Adversarial

By definition, the existence of an intelligent adversary is inherent to cyber security.

There are numerous definitions of *cyber security* and they often justify this assertion implicitly. For example, the definition "Cyber security refers generally to the ability to control access to the networked systems and the information they contain," [6, pg.1] implies the existence of something capable of accessing the systems and information against our wishes. Likewise, "Security is a composite of the attributes of confidentiality, integrity, and availability" [4, pg.13] implies an entity which could break that confidentiality, integrity, and availability. Section 3.3 and Section 4 describe reasons for relying on foundational definitions in this way, and some of the dangers of doing so. For now, of the three properties we will spend the most time on the adversarial, as it is foundational.

To explore the adversarial property, we will use as a point of reference the work of contemporary science historian Peter Galison,

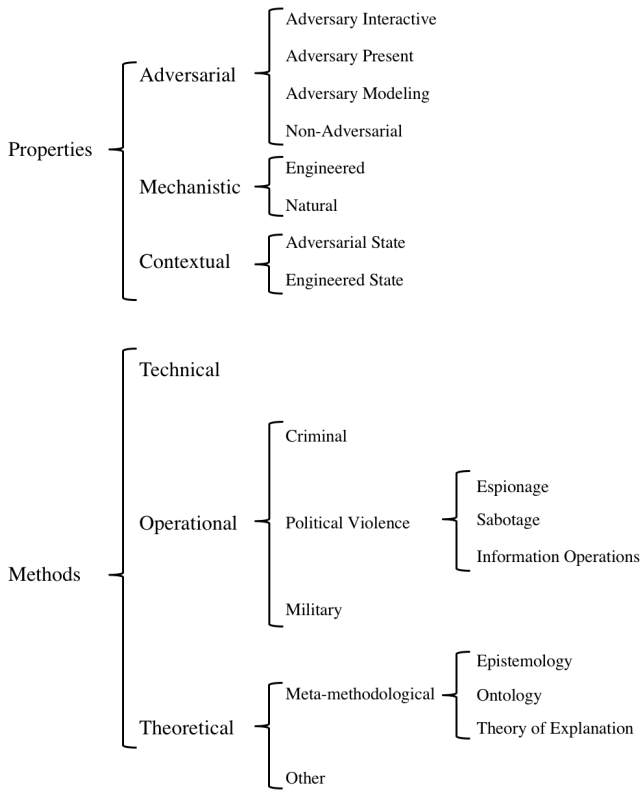

Figure 1: A Taxonomy of Characteristics of Cyber Security

who used several theoretical ideas from post-World War 2 control-theorist[1] Norbert Wiener. Galison presented his ideas to the JASON government advisory group for their 2010 report on 'Science of Cyber-Security' [38], and also as a keynote to a 2012 science of security workshop [25].

Galison adapted terminology from Wiener to identify an **adversarial science**[2] as one that involves an "active oppositional intelligence" [24, pg.232] who in Wiener's words "is determined on victory and will use any trick of craftiness or dissimulation to obtain this victory" [24, pg.231]. Galison suggests cyber security is one such adversarial science [25], along with game theory, control theory, and military operations research [24, pg.232]. In contrast, **non-adversarial sciences** such as Biology and Chemistry involve overcoming "chance and disorder" [24, pg.231] including a mere lack of knowledge.

We expand upon their basic distinction to consider three descriptions of adversarial disciplines. A discipline practitioner can *interact* with their adversary, each trying to disrupt the other's mission[3], but a practitioner can also work to accomplish their mission in the *presence* of an adversary without directly attacking their adversary's mission, or merely *theorize* about adversaries without the theorizing being part of an adversarial interaction.

First, the term **adversary interactive** describes a discipline in which the discipline practitioner and their adversary each attempt to disrupt the other's mission.

An example from cyber security is in attack-defense "capture the flag" exercises. The nature of these events vary, but a typical one is described by Cowan et al. [12]: participating teams share a computer network and score points by maintaining a server that provides a service which is known to be vulnerable to exploitation. Participating teams can also score points by exploiting that service while it is being run on their adversaries' servers. So all participating teams are adversaries of each other.

A non-cyber example of adversary interaction is in "active defense" criminal trials, in which defending and prosecuting lawyers each attempt to establish narratives regarding the defendant while attacking the validity of the other lawyer's narrative.

Second, the term **adversary present** describes a discipline in which an adversary attempts to disrupt a discipline practitioner's mission. The practitioner does not attack their adversary.

As an example from cyber security, the JASON group ultimately understood cyber security to be "science in the presence of adversaries" [38, pg 15], a formulation that seems to be shaped by cryptography [38, pg 2], which is about "communication in the presence of adversaries" [62, §2]. For example, communications that aspire to *privacy* define an adversary and their capabilities. The adversary may observe encrypted messages, or observe example encrypted messages with their unencrypted versions, or the adversary may have access to a version of the cryptographic system [62, §6.3.1]. The goals of protecting against the adversary can then be used to produce a formal and provable characterization of security. The issue of importance here is that the adversary is attacking the discipline practitioner's privacy, but the practitioner is not attacking the adversary back.

As a non-cyber security example, in academic peer review ideas are "attacked" to seek flaws, and while the scholar being reviewed may provide defenses and justifications, they are generally not expected to attack the reviewers back.

Third, an **adversary modeling** discipline involves developing a representation or theory about an adversarial relationship. The theorist does not participate in that relationship.

As a non-cyber example, consider an ecologist studying lynx-hare population dynamics in the wild [57]. Clearly such a predator-prey system is adversarial, but while the lynxes and hares are adversaries of each other, they are not adversaries of the ecologist.

As a cyber example, consider Nguyen et al.'s application of Stackelberg games to security problems [52]. This work is motivated by particular specified use cases, but the focus is on developing general methods and principles. At one extreme, game theoretic efforts such as von Neumann's seminal paper [79] may have no identifiable applications to a specific adversary. At the other extreme, even theoretical papers can be used in adversary interactive contexts. We need a way to distinguish between adversary modeling and adversary interactive descriptions.

---

[1] Wiener self-identified as a cyberneticist, referring to the science of control, which he founded. We use the term 'control theory' to avoid confusion between cybernetics and cyber security. In prior work we discuss the importance of control in characterizing adversarial cyber security interactions [63].

[2] Wiener and Galison use the terms 'Manichean' for adversarial and 'Augustinian' for non-adversarial, which resonate aesthetically but which potentially obscure their meanings.

[3] Where a **mission** is a task that is carried out during an **operation** [15].

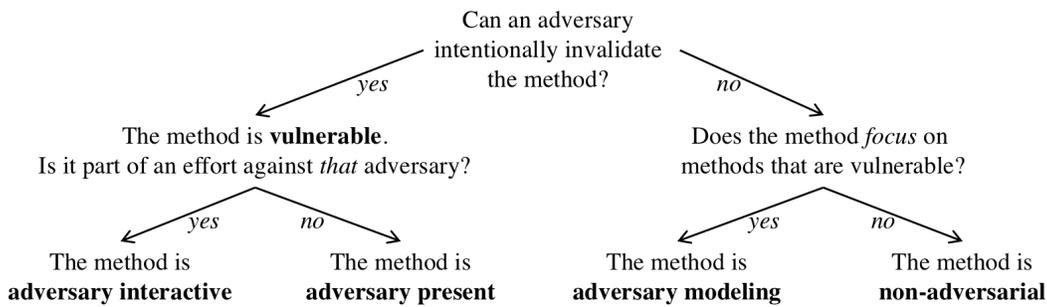

Figure 2: Characterizing Methods by Invalidation

So we have defined several adversarial descriptions. But how can we determine the right description for a method?

We begin by considering how Wiener describes adversarial methods:

> The distinction between the passive resistance of nature and the active resistance of an opponent suggests a distinction between the research scientist and the warrior or the game player. The research physicist has all the time in the world to carry out his experiments, and he need not fear that nature will in time discover his tricks and method and change her policy. [81, pg 36]

In Wiener's non-adversarial setting, nature is not attempting to disrupt the research physicist's mission, so the physicist's task is easier. They have "all the time in the world" to carry out their experiments and, presumably, to validate their experimental methods. But in Wiener's adversarial setting, the adversary will undermine the methods of the warriors and game players. This adversary will do so by observing the "tricks and methods" used and intentionally invalidating them.

We call this the **vulnerability property of methods**: the extent to which an adversary can intentionally invalidate a method's ability to support a mission, for a defined perspective and scope. We can use this to determine an adversarial description, as shown in Figure 2. First, we define a *vulnerability criterion*: is the method vulnerable to an adversary?

If the vulnerability criterion is met, we then ask if the method is part of an effort to disrupt the mission of *that* adversary who is invalidating it. If so, then the method is described as adversary interactive; if not, then it is adversary present.

If the vulnerability criterion is not met, then we ask if the subject of the method's effort is an adversarial interaction. If so, then the method is described as adversary modeling; if not, then it is non-adversarial.

The vulnerability property is defined "for a defined perspective and scope" because adversarial descriptions are relative to their circumstance of use.

In terms of *perspective*, consider a terrorist who mails explosive devices to university scholars. Any scholar could become the target of such an attacker, but it does not follow that all scholarly methods are inherently adversarial. Rather, this introduces the **subjective** aspect of adversariality: that an adversarial characteristic is limited by perspective. From the terrorist's perspective, the scholars may indeed all be part of a political or cultural conflict. From the scholars' perspectives, this may not be the case. Therefore there are at least two subjective interpretations of the adversarial situation. The terrorist's perspective may seem more comprehensive, but consider a third-party intelligence operative who is observing this activity, unbeknown to the terrorist and scholar. This third party's perspective is even more comprehensive.

In terms of *scope*, consider a truck that is speeding through a combat zone, occasionally swerving to avoid any possible enemies. An enemy aircraft is tracking the truck unbeknown to it, waiting for the right shot. In this encounter-level scope, the truck is adversary present. But if the truck is carrying munitions to an anti-aircraft weapon that is tracking the enemy aircraft, then in this operation-level scope the truck is adversary interactive.

We conclude this subsection with some additional examples.

*Example 1: adversarial shift.* Consider an ecologist studying lynx-hare interactions for purposes of maintaining stable populations in a wildlife preserve. In this ecologist's context, their studies are adversary modeling. If a cyber security researcher read the ecologist's studies and applied its theoretical insights to an active cyber conflict, then this use would be adversary interactive. The opponent targeted by this method might also read the ecologist's studies and change their policy to invalidate their use, but this would only invalidate the ecological studies' adversarial application, not their original use. This is because the opponent is only targeting the adversarial adaptation of ecological methods, not the original ecological use of the methods.

Fancifully, the opponent may even decide to disrupt the ecologist's studies, perhaps secretly corrupting the ecologist's data to ultimately mislead the security researcher. Such a disruption would be in the scope of the adversarial use of the ecological methods. The ecologist, focused on their scope of lynxes and hares, may nevertheless be confused by the unexpected data. This confusion is due to their subjective perspective, and is a result of the **adversarial shift** that occurred when their methods moved from a non-adversarial to an adversary interactive use. The security researcher might thus inform the ecologist that their methods are being used adversarially, and that *validating a vulnerable method involves considering its invalidation by an intelligent adversary.*

At this point, the ecologist could reject the cyber security context, and try to return to the scope of their adversary modeling studies. But even then, they would be wise to change their methodology to protect their data from any further interference, in the same way that they might protect their credit card data for example. A well-constructed methodology in this case would be adversary present, since it would be on guard against any possible adversary. Alternately, the ecologist might become enraged by the threat to their research and dedicate themselves to the conflict, focusing their efforts not only on protecting their data but on producing studies that assist the cyber security researcher in their conflict. In this case, the ecologist would have adversary interactive methods.

*Example 2: intentionality and the vulnerability property.* A botanist's studies of flora in a wildlife park are adversary present, because to complete their studies the botanists need to maintain their safety in the presence of predators such as tigers. The botanist may also need to maintain their safety during any floods that may happen, but the floods will not invalidate their efforts intentionally in the same way a tiger will. If the botanist stays within a radius of high ground and shelter, the flood will not seek some way of overcoming this. But if the botanist stays within a radius of a safe truck, the tiger may learn to stealthily creep between the botanist and the truck before attacking. Likewise, a natural disaster will not damage a network as methodically as a malicious human attacker can.

This is the first type of intentionality: the purposeful, reactive nature of the attack, which might be formally represented with a Belief-Desire-Intention [68, §4.2.1] or similar model. This type of intentionality is part of what makes the adversary *intelligent*, and is part of the vulnerability property of methods.

The second type of intentionality relates to whether the tiger *intends* to stop the botanist from examining the flora. In other words, was the vulnerable mission in question – the study of the flower in this case – the intended target of invalidation? In fact, the tiger attacks the botanist out of hunger rather than a fierce determination to save the flora from examination. The tiger does not care about the botanist's mission, any more than a network intruder cares about the many missions of people whose traffic they are sniffing (and confidentiality they are violating) while searching for a particular target. This second type of intentionality is not a necessary part of the vulnerability property of methods, although it characterizes the adversarial relationship in terms of the particular missions that are in conflict.

*Example 3: subjective effects on adversarial characterizations.* Consider a network intruder who is illicitly sniffing traffic for financial information. If a legitimate user of that network decides to encrypt their email out of a nonspecific wish for privacy, they are not attacking the mission of *that specific intruder*, but rather, of any such intruder. Likewise, that intruder is not attacking the confidentiality-preserving mission of that specific user, but of any such user. So they are in an adversary present relationship with each other.

If the intruder decides to target that user and bypasses their encryption implementation by installing a keylogger on the user's workstation, then the intruder has entered an adversary interactive relation with the user, while the user is still adversary present.

The intruder may assume that the network's administrator is logging various types of activity and scanning various files for malware. The intruder and the admin are attacking the missions of *any such* adversary. So they are in an adversary present relationship with each other.

Imagine the admin detects the keylogger and begins to carefully monitor their network traffic. The admin is reacting to the intruder's actions, so the admin is adversary interactive with the intruder, though the intruder is still adversary present.

The admin notifies the user about the keylogger, so the user begins using a different workstation and email account; that user is now also adversary interactive, because they are reacting to that specific intruder's actions.

Eventually the admin and user may believe the intruder's software has been removed from their machines, so they once again assume they are adversary present. The intruder, having noticed the keylogger removal and the email account change, is now adversary interactive and is planning their next move.

In this way the adversarial relationship iteratively changes, and the engineered context changes with it. The next two sections describe this.

## 2.2 Mechanistic

Cyber security involves both engineered and natural aspects, but is dominated by the engineered.

In this context 'engineered' and 'natural' are defined in terms of the *mechanistic approach* in philosophy of science, as applied to computer security by Hatleback and Spring [34] [70].

Mechanisms are organized entities and activities that produce regular changes in a process [34, pg 443] [47, cited]; a definition that supports a philosophical perspective on the sciences. Hatleback and Spring use this to distinguish between engineered and natural mechanisms in computer science, and in computer security. **Engineered** mechanisms are artificial, such as DNS schemes and DNS cache poisoning attacks, and **natural** mechanisms relate to the physical world (and are thus also called *physical* mechanisms), such as the cognitive aspects of human-computer interaction [34, pg 444-446]. This is contrasted with an *engineering vs. scientific* distinction [65], which precludes the possibility of considering a scientific approach to its engineering aspect and which lacks the clear connection to the mechanisms involved [34, pg 445-446]. A similar argument can be made against distinctions such as *research vs. development* [3] or *applied research vs. basic research* [14].

The engineered mechanisms are those which can be changed [34, pg 445], and both sides of the adversarial interaction will change the engineered environment to further their mission and impede the other's mission.

For cyber security, the **engineered environment** is the set of all implemented mechanisms. It is constantly changing because of the security struggle as well as due to the evolution of technology. A struggle takes place in the engineered environment, each participant seeking to gain control of the environment and to modify it to their benefit. Although the natural mechanisms cannot be changed, they act as constraints in which the struggle takes place, and the adversary that understands them best will be best positioned to maneuver around them.

Generally speaking, if two participants are in conflict in an engineered environment, one participant will modify the environment until it is suitable enough for their needs, at which point the other participant will do the same, and so on. This creates an endless **adversarial cycle**. In cyber security this has been called an "arms race" [17] [53] [72], a "cat and mouse game" [18] [23], or a game of "whack-a-mole" [2] [13] [66]. The cyber security methods that are relevant to one iteration of the cycle may not be relevant to the following cycle. In fact, a competent adversary will ensure that this is the case.

## 2.3 Contextual

Cyber security is **contextual**: characterized by the state of the adversarial conflict, and the state of the engineered environment, for a particular time and place.

The context is defined by the *adversarial state*: what each participant is doing, what their plans are, what they hope to achieve, and by the *engineering state*: what the combatants are struggling for control over, who currently controls what aspect of the environment, and how much each side understands of the situation.

Because this context is adversarial, its exact state cannot be known with certainty. Consider a security practitioner who wants to know if a USB drive contains malware. Perhaps they study it with a software tool. But how do they know if the tool, or the operating system it runs on, is not infected in such a way as to deceive them? Perhaps they install the operating system directly from the manufacturer, and compile the tool from source code without connecting to the internet. But how do they know if the compiler, operating system, or even hardware were not compromised with malware by the manufacturer, by an insider threat perhaps? Even if the practitioner were able to construct their own hardware from sand and plastic and write their own tool in binary, they still might make a design or programming mistake enabling a malicious exploit. Of course practically speaking at some point the practitioner is satisfied enough with the relative safety of their tools to proceed with their analysis, but the fact remains that *their sense of security has no absolute foundation*. In Section 3.3 we will see similar epistemological regress problems.

Not only is context changed by the adversary, but it also evolves under the pressure of consumer demand, business product creation, software and hardware innovations by entrepreneurs, and so on. So the context changes in unpredictable ways. As time goes on, the amount of change increases. And a method that was valid at one point in time may not be valid in another. That likelihood of contextual invalidation increases as time goes on.

To consider ways of managing this, we will consider the types of cyber security methods involved.

## 3 METHODS IN CYBER SECURITY

For our focus on validating methods, we will consider the three types of methods involved in cyber security shown in Figure 1: Technical, Operational, and Theoretical.

Because the engineered environment of any particular problem varies, one set of methods relates to solving problems related to the current engineered environment. These are the *technical methods*.

And because the adversary and the state of the adversarial struggle vary, one set of methods relates to deploying those technical resources to achieve goals in the current adversarial context. These are the *operational methods*.

Finally, confronted by seemingly endless adversarial cycles, cyber security practitioners occasionally see the need to systematize their discipline or to generalize problems across contexts. So one set of methods relates to exploring the theory and the science of the discipline. These are the *theoretical methods*.

In Section 3.1 we describe ways of validating technical methods. This includes considering operational invalidation, so we describe operational methods in Section 3.2. However, the technical validation methods are meta-methods, which themselves need to be validated. Some basic concepts to do so are described in Section 3.3, and validating such meta-methods is described in Section 4.

## 3.1 Technical Methods and their Validation

We need not develop a taxonomy of technical methods because such taxonomies already exist in the form of the ACS Computing Classification System, for example. Instead, we will note that the continually changing engineered environment provides an endless source of technical problems to be solved by security practitioners. Although this creates an exciting amount of interesting work, it is also inherently problematic: having implemented a technical method and having tested for **technical invalidation** (e.g. by performing unit tests, processing sample data, etc.) the security practitioner is often encouraged to hurry on to the next technical problem, without systematically validating the method.

This includes validating against **adversarial invalidation**: that a method has been undermined or counteracted by the antagonist and is no longer as useful, as described in Section 2.1. Beyond the efforts of an adversary, **contextual invalidation** can also occur when natural technological innovation changes the engineered environment in a way that the technical method is no longer usable, as described in Section 2.3.

A final type of danger is indicated by our use of the term "invalidation." The vulnerability property in Section 2.1 referred to the invalidation of a method's ability "to support a mission." A method that is useful for one mission (and the operation that it is a part of) may not be useful for another. So **operational invalidation** occurs when a mission is not conducive to the effective use of a particular method.

Table 1: Validating Technical Methods

| Invalidation | Validation |
|---|---|
| technical | varies per sub-discipline |
| adversarial | identify, detect, react |
| contextual | identify, detect, react |
| operational | |

We have identified a set of potential invalidations, but now we cannot provide a normative and universal set of validations, for several reasons. First, the properties of cyber security guarantee that such a set of validations would themselves risk adversarial or contextual invalidation. Second, the details of validating, for example, an adversary present cryptographic method in one era are likely to be much different from the details of validating an adversary interactive network defense method in another era. And third, exploring the many different types of validation is a task that should be performed on solid foundations, which this paper is part of the process of establishing.

As a point of contrast, technical validations (i.e. demonstrations that the engineered aspects of the method work as designed) vary by method and sub-discipline because they are more established: in some cases they require a proof and in other cases a user study, for example. Adversarial, contextual, and operational validations are not as established for the reasons described in the previous paragraph. So instead we will specify what a validation should accomplish. The validation should *identify* the ways in which the invalidation could happen: for example, by determining what contextual developments invalidate the technical method's assumptions, or by specifying the operational parameters (such as assumptions, use cases, and potential misuses) required by the method. The validation should describe ways to *detect* when such an invalidation has actually occurred, in terms of sensors or logs, for example. The validation should also describe how to *react* upon detecting an invalidation: how is damage mitigated? What is the next iteration of the adversarial cycle?

## 3.2 Operational Methods

The previous section argued that technical methods in cyber security are ultimately used *to support operations*.[4] So we should be specific about the types of operations we are talking about. We will use a taxonomy derived from Rid [61] which as shown in Figure 1 includes Criminal, Political Violence, and Military categories. We will consider methods related to all participants in the adversarial struggle: for example, both criminal methods and law enforcement methods. Of course we do not condone criminal methods, but to understand the conflict we must understand the crime, and be able to describe the threat.

*3.2.1 Criminal.* This category of operation is the one most people are likely to encounter, as it involves illegal activity between individuals and organizations, not nations or insurgent groups which wish to defeat nations. Political violence or military activity may break international norms or even international law, and therefore be considered crimes, but for our purposes we will consider an activity *criminal* if it is breaking a nation's laws and it primarily involves victims within that nation.

Criminal methods include ransomware, identity theft, and other cases of stealing from businesses [76] and individuals. Methods against criminal activities include "computer safety" approaches, as well as security procedures both by organizations and by home computer users [36]. This can range from help desk scripts (which might begin with, for example, "did you try rebooting?") to government guidelines and suggestions for businesses [74], to digital forensic methods used when bringing criminals to trial [10].

---

[4]This is unlike technical methods in non-adversarial sciences, which are used *to develop theories* – to provide evidence for theories, or to falsify theories, or to compare theories, depending on the epistemological commitments of the practitioners. See Section 3.3.

*3.2.2 Political Violence.* This involves conflict between nations or insurgent groups, as opposed to individuals or organizations. Political violence includes espionage, sabotage, and information operations.

The goal of cyber *espionage* is "to capture data of the opposing force" and includes "exfiltration, monitoring, and theft of digital information." [69, pg 114-115] These operations may be "advanced persistent threats" by highly-skilled teams whose work is difficult to detect [48].

Cyber *sabotage* operations involve, for example, attacks on national infrastructure such as industrial control systems and power grids [69, ch. 11-12].

Finally, the understanding of *information operations* varies somewhat by country [28]. In the U.S. its purpose is to "influence, disrupt, corrupt, or usurp the decision-making of adversaries" [15, pg. 115] but is sometimes known as "influence campaigns" [43], "strategic messaging" [39, pg. 24], or "information warfare" [31]. Information Operations have been carried out by Russia [55] based on well-developed theories [22], as well as by China [8] [11, pgs 23-25], Israel and their antagonists Hezbollah and Hamas [69, Chapter 4], and ISIS [40].

*3.2.3 Military.* As explored by Rattray and Healey [60], these include situations that overlap with political violence as well as those that purely support kinetic action. The U.S. military produces a large number of documents on operational methods, and also has a strong tradition in developing technical methods to support operations as well as for acquisitions [27] [58]; these methods are gradually being adapted to cyber security [26] [46].

## 3.3 Theoretical Methods

Cyber security as a field is by necessity very focused on its immediate context. Even so, practitioners occasionally ask: how good of a job am I doing at this? and: will I always just be "putting out fires"?

In recent years, security researchers have sought such theoretical methods, hoping to find fundamental principles to drive innovation and to predict the effectiveness of technical tools and operational policies [45][50][67]. Security researchers have looked to physical and social sciences for universal laws and considered whether these laws will follow rather than inspire security solutions [19][41]. Several studies suggest improvements to methodology [51][64] and scientific reporting [9], and describe the nature of scientific claims [33] [75], but no consensus has yet developed [35].

This paper is motivated by these types of questions. As a point of approach, we are considering the validation of methods. In colloquial terms, we may ask: how do I know if my security methods are any good? More formally, a **meta-methodology** is an "investigation into principles of method and their justification" [54, pg. 80]. We use the term *validation* as a synonym for justification. Section 3.1 included one such validation, summarized in adversary present 1.

Every method has a meta-methodology; if it is completely unexplored, it may be called "more art than science," "a black magic," "artisanal," or "tradecraft," suggesting that the method is validated by an expert on a case-by-case basis, usually after trial-and-error.

A meta-methodology[5] includes an *epistemology*, an *ontology*, and a theory of *explanation*.

**Epistemology** involves "the nature and justification of valid knowledge [and] the procedures by which models or explanations are constructed." [7, pg. 54] As an example of why this might be useful, consider that to validate a methodology we seek a meta-methodology. But this leads us to ask whether to validate a meta-methodology we should seek a meta-meta-methodology. Which in turn requires a meta-meta-meta-methodology and so on. Informal discourse and popular science often use the imagery of "turtles all the way down" [82] but in philosophy this is an epistemological *regress problem* [71, §3.3], which has been addressed since the time of ancient Greece. Resolutions include *foundationalism*, in which some beliefs are assumed (e.g. as with mathematical axioms); *coherentism*, which requires that assumed beliefs be part of a sufficiently-comprehensive system of beliefs, and *infinitism*, which accepts the infinite regress [21].

Additional epistemological questions include debates on the existence of a universal scientific method, and whether such a method should allow induction, or focus on validation or falsification [1].

**Ontology** "seeks to describe the basic categories and relationships of existence thereby providing an account of the types of things there are and the manner (or mode) of their being" [7, pg. 54]. For example, this paper makes ontological commitments regarding the properties of cyber security and has explicitly characterized adversarial interactions with a decision tree.

**Explanation** is a way of making "claims that go beyond our experiences and purport to discover underlying causes." [7, pg. 54] This is the aspect that creates understanding and allows prediction. In cyber security, this is difficult due to the vulnerability property of methods: any explanation is subject to invalidation by an adversary or by the evolving engineered environment.

The next section will summarize the challenges faced by cyber security meta-methodologies, and suggest ways of managing those challenges.

## 4 VALIDATING META-METHODS

For most security practitioners, the validation of technical methods in Section 3.1 will suffice. But to be comprehensive, we also need to consider *the validation of meta-methods*.

In cyber security, meta-methods are also vulnerable to adversarial and contextual invalidation. An adversary who learns of a given meta-method can in principle seek to modify the engineered environment, and can change their operational plans to invalidate that meta-method. Worse, at any given moment a security practitioner cannot be sure of the extent to which an adversary has already invalidated their methods and meta-methods.

For example, imagine that a researcher builds a system that measures the security of a system. The researcher wonders how they can be sure that their system works well. In other words, their system is a technical method, and they want a theoretical meta-method for validating their system.

Imagine that the researcher decides to use data to test the system, so they collect data from operational systems. Immediately they are confronted with a number of questions: how do they find a "ground truth" about whether the data they collected is in fact secure or not? How do they know if the engineered environment will change in ways that make the data obsolete? How do they know if an adversary will discover their system, and change the engineered environment or their behavior in a way that will circumvent the system without the researcher realizing it? Because if the adversary does so, then they have invalidated the researcher's meta-method.

So, simply by asking "how do I know if it works well," the researcher has found themselves in a dizzying theoretical environment, assailed by adversarial invalidation, on the precipice of a regression (what meta-method should validate their meta-methods, etc.), gingerly but inevitably heading towards a never-ending adversarial cycle.

But now we have surveyed the properties and the methods of cyber security. So consider the validation of technical methods described in Section 3.1 and summarized in Table 1. That validation is a meta-method, and thus itself vulnerable to invalidation, or a regress of meta-methods. But now we have three well-studied epistemological approaches to resolving this regress. And we have the operational methods of Section 3.2, each with its own tradition, with more or less technological sophistication, of validating methods and making decisions in adversarial environments. And indeed, we can seek further guidance in these adversarial disciplines.

For example, law enforcement *forensic sciences* are familiar with criminal efforts to invalidate their methods, and have managed questions about the validity of new digital forensics techniques [10] [37].

Another example is *strategic theory*, which considers war and political violence, as well as the national security impacts of crime [5]. It considers conflict that is adversarial by definition, engineered [30, pgs. 38-40], and contextual [32, pg 43]. It is built on a substantial history of operations and technical methods [58] [27], which serve as material for debates on whether there are universal laws and methods of conflict [32] [56] [77].

We are not saying that cyber security should be addressed purely from the perspective of strategic theory, the forensic sciences, or any other operational discipline. But security is interdisciplinary, and we can productively look across disciplines for methodological inspiration. In Section 4.1 we will give an example of this, using Clausewitzian critical analysis from strategic theory[6]. After that in Section 4.2 we will describe some of the characteristics of this interdisciplinary methodological endeavor.

### 4.1 Critical Analysis as a Performative Meta-Method

Clausewitz considered the relation between critical analysis, theory, and war[7]. His basic argument is that while it may be possible to have a doctrine for how to "organize, plan, and conduct an engagement," we cannot come up with a static set of rules or doctrine to conduct war in general, across time. Instead, theory should study "ends and means" and their related factors. The way this is done is through critical analysis, where the word *critical* is used in the sense of *critique* rather than *criticism*.

---

[5]This formulation is adapted from Bevir's [7] consideration of the meta-methodology of political science. We do not mean to suggest that using these terms should be mandatory for security professionalism. Rather, these terms are born from well-studied disciplines that can be helpful in resolving difficult questions in theoretical methods.

[6]In strategic theory texts it has been noted that "because Clausewitz said something did not necessarily make it true, but it did make it worth considering." [5, pgs.13].
[7]Unless otherwise stated, the description and quotes in this subsection are taken from Book 2 Chapters 2-6 of *On War* [78].

Clausewitz describes critical analysis as having three parts.
(1) Identification of *facts*
(2) Determination of *causes* of the facts
(3) Investigation of the *means* used and the *intentions* behind their use

We will explore this in terms of the parts of a meta-methodology as described in Section 3.3.

*Epistemologically*, knowledge is constructed through a study of past incidents in part 1, and continuing through and analysis of "relationships between phenomena" in parts 2 and 3. The analyst also has to consider "what might have happened" as well as what actually happened.

Determining causality in the second part can be extremely challenging so there are bound to be gaps, but "all a theory demands is that investigation should be resolutely carried on till such a gap is reached. At that point, judgment has to be suspended."

Clausewitz discusses several additional challenges in determining causes throughout the process. In describing how analysis of causes and intentions work together,

> "we can follow a chain of sequential objectives until we reach that one that requires no justification, because its necessity is self-evident. In many cases, particularly those involving great and decisive actions, the analysis must extend to the ultimate objective, which is to bring about peace."

In this way Clausewitz resolves the regress foundationally.

The second and third parts of the process require a "working theory" to be useful, but the aim of the analysis should not be a "mechanical application of theory." Instead, the theory should be tools to help the analyst in their judgments. If the analyst is studying a conflict and one of the combatants acts contrary to the "principles, rules, and methods" of the working theory, the analyst should investigate the reasons for this, and the analyst "has no right to appeal to theoretical principles unless these reasons are inadequate." If that contrary behavior ends in failure, the analyst should not automatically assume that the unexpected behavior was what caused the failure. Likewise, if that contrary behavior ends in success, the analyst should not assume that the working theory was incorrect. The role of the theory is not to provide "laws and standards" but "findings" that assist in the analyst's "judgment."

*Ontologically*, the analysis is made of historical facts (individuals, units, events, actions), the causal chains of those events, and a description of those events in terms of the means that the belligerents used, and the intentions behind those selected means. Furthermore, they should be described in terms of operational language: "the language of criticism should have the same character as thinking must have in wars," so that the analysis retain its practical value.

The theory of *explanation* is provided by the final analysis of causes and means which the analyst has constructed.

However, producing the explanatory analysis is not the only product of this process.

> "Critical analysis being the application of theoretical truths to actual events, it not only reduces the gap between the two but also accustoms the mind to these truths through their repeated application."

Not only is theory changed by the critical analysis, but it is meant to prepare the analyst for future, unexpected contexts. "[Theory] is meant to educate the mind of the future commander, or, more accurately, to guide him in his self-education, not to accompany him to the battlefield." Which is not to claim that all guidelines should be rejected: it is acceptable if "the theorist's studies automatically result in principles and rules," but theory is "never to construct an algebraic formula for use on the battlefield. Even these principles and rules are intended to provide a thinking man with a frame of reference... rather than to serve as a guide."

In other words, critical analysis is **performative**: the fact that the analyst is performing it is as important as the conclusions they reach, and performing it during an action is its own validation. After all, any pre-action theoretical conclusions may be invalidated by the adversary in the next operation. Because adversarial operations are so context-dependent, rather than simply learning rules about what to do, the combatant should learn how to solve problems in real-time, with theory to provide a guiding "frame of reference." Critical analysis gives them practice in the skills of producing causal explanations, questioning theories when needed. This is what Clausewitz calls "creative ability" in war.

To describe this in terms we have introduced previously, the methods related to the "engagement" described at the beginning of this subsection are analogous to technical methods in cyber security: when related to "material factors," this type of military creativity is less necessary. But there is a level at which "the intellect alone is decisive," and methods at this level are validated through critical analysis: before the action as training, during the action as decision-making, and after the action as preparation for the next event.

## 4.2 Interdisciplinary Generative Meta-Methods

It would be self-contradictory to offer a single way of validating meta-methods. We have described meta-methodology as vulnerable to invalidation by adversarial action or context. We therefore benefit from **generative** meta-methods: ways of developing additional methods.

Consider the demands that we place on meta-methods. They must validate our methods while themselves avoiding invalidation, they must manage regress problems, and they should avoid adversarial cycles. Ideally they should work universally, but because of the cyber security's contextual property we will usually need to be satisfied with a way of telling when the theoretical method becomes invalidated or obsolete.

We have earlier described security as interdisciplinary, so we will once again seek inspiration across disciplines. This paper has already made use of philosophy, which is one of the humanities. Consider: in many contemporary views, humanist meaning does not equal objective truth but represents an attempt to explore, to "problematize", to create thought experiments, etc. In a way, this is chasing the infinite regress: asking questions, which are then questioned, which are themselves questioned, and so on. (Just as this characterization of the humanities would itself be questioned, which would be questioned, and so on.) After the popularization of the internet, the humanities initially focused on applying engineering approaches to the humanities, by applying techniques from natural language processing to texts. But in recent years it has also applied humanities approaches to engineering [16, §1.2] [59, ch.1] where the purpose is once again to ask good questions, not seek universal answers[8]. This will be our inspiration.

Adversarial struggles require solutions to unique and unforeseen circumstances. Good methods account for this and generate these solutions. Good meta-methods allow for the generation of the appropriate meta-methods. Adversaries compete to do this best. Even in a lull, researchers develop methods and theories, preparing for those unforeseen conflicts. Developing skill in their analyses involves discussion with other researchers not just to transmit facts, but also to judge the value of the way one is thinking analytically, and learning new ways of thinking, strengthening conceptual abilities for the next adversarial cycle. When conflict occurs, researchers study the effectiveness of their methods. They also question the effectiveness of their studies, in disentangling a method's effect on the outcome, for example. Adversaries compete to do this best as well. And so on.

This may seem overly abstract, so let's consider an example.

Imagine two nations, near-peer rival powers. The cyber security researchers of these nations develop technical solutions while stopping online crimes, defending against industrial espionage, securing critical infrastructure, and so on. They talk amongst and between themselves, and with other nations, sharing various methods and theories, and not sharing others.

Imagine these two nations become involved in an overt force-on-force cyber-only conflict [60, pg.88,92-95]. The engineered environment, distorted by the conflict and having evolved under the influence of millions of consumers, is unlike that of any year past. The combatant nations' researchers apply their technical methods in a variety of operations. Their ability to generate effective tactics, to invalidate their adversary's tactics, to plan effective operations and meta-methodologically estimate their chance of success, etc. are all enabled by the quality of their theories and the performative abilities of their security personnel.

The conflict ends. The rival nations and others examine the conflict, identifying what happened and what caused it to happen that way. Various theories are proposed, those are questioned, which are themselves questioned and serve for new theories, and so on.

This example is highly simplified. Realistically, such a scenario would also involve the influence of activists, international businesses, criminals, knowledgable civilians, and so on, each reacting in their own ways. However, it highlights the interaction between technical, operational, and theoretical methods.

Practically speaking, this type of theorizing can be made part of operations research, the first step of which is to perform an analysis defining the problem and constructing a model before applying the appropriate method [73, pg.8-9]. Wargaming, optimization, capability assessment are all a part of this. At this stage, when the methods are stated, so can a validation of the methods. The purpose is not to provide unarguable answers. The purpose is to provide possibilities, to support decision-making.

## 5 SUMMARY

In this paper we make several contributions relevant to the foundational problem of validating methods in cyber security.

First, we provide precise definitions of cyber security's adversarial, mechanistic, and contextual properties. This includes a decision tree to determine the adversarial characteristic of a method. This now enables us to unambiguously discuss, for example, the difference between methods in adversarial and non-adversarial disciplines, as well as phenomena such as adversarial shifts and adversarial cycles.

Second, we build on these definitions to describe three sets of methods in cyber security: technical, operational, and theoretical. Cyber security is inherently operational, so we describe various types of cyber activities in terms of operations. We then describe a well-considered theoretical meta-methodology. This now enables us to state, for any technical method, what the validation needs to accomplish: it should identify, detect, and react to potential adversarial, contextual, and operational invalidations.

Third, we describe ways of validating theoretical methods. We describe critical analytic and generative meta-methods, and provide a scenario showing the role that theoretical methods would play in an extended cyber conflict. This provides us with at least two possible ways of further studying foundational principles.

The concepts we define are intuitive and some have been alluded to in the literature. But they have not until now been precisely defined in the form of a taxonomy and grounded in mature disciplines. Because of the contextual nature of cyber security, we cannot provide a thorough listing of guaranteed ways of validating methods. Instead we have provided a framework spanning from first principles to validation requirements to theoretical considerations, for use by researchers and professionals in their given context. It turns out that the best way to validate a method is not an algorithm, a proof, or a technology. The best way to validate a method is people.

## ACKNOWLEDGMENTS

Thanks to Kevin Bush, Melva James, and Eric Hatleback for their insightful comments, to Jeremy Blackthorne for relevant conversations, and to John Wilkinson for his support and encouragement.

This material is based upon work supported under Air Force Contract No. FA8721-05-C-0002 and/or FA8702-15-D-0001. Any opinions, findings, conclusions or recommendations expressed in this material are those of the author(s) and do not necessarily reflect the views of the U.S. Air Force.

Distribution Statement A. Approved for public release: distribution unlimited.

## REFERENCES
[1] Peter Achinstein. 1998. Demarcation Problem. In *Routledge Encyclopedia of Philosophy*. Routledge.
[2] Mohammed H Almeshekah and Eugene H Spafford. 2014. Planning and Integrating Deception into Computer Security Defenses. In *Proceedings of NSPW*.
[3] American Association for the Advancement of Science. [n. d.]. Definitions of Key Terms. https://www.aaas.org/page/definitions-key-terms Last checked August 21, 2017. ([n. d.]).

---
[8] Of course in doing so, humanist discussion produces material that may be useful in a given context. Philosophy in particular is similar to "pure math," which explores relations and produces concepts and conclusions that may be useful in operational or scientific contexts, but whose main motivation is exploration, not utility [20].

[4] Algirdas Avizienis, Jean-Claude Laprie, Brian Randell, and Carl Landwehr. 2004. Basic Concepts and Taxonomy of Dependable and Secure Computing. *IEEE Transactions on Dependable and Secure Computing* (2004).

[5] J Boone Bartholomees Jr. 2012. A survey of the Theory of Strategy. In *U.S. Army War College Guide to National Security Issues, Volume 1: Theory of War and Strategy*. U.S. Army War College, Carlisle Barracks, PA.

[6] Jennifer L Bayuk, Jason Healey, Paul Rohmeyer, Marcus H Sachs, Jeffrey Schmidt, and Joseph Weiss. 2012. *Cyber Security Policy Guidebook*. John Wiley & Sons.

[7] Mark Bevir. 2008. MetaâĂŘMethodology: Clearing the Underbrush. In *The Oxford Handbook of Political Methodology*. Oxford University Press.

[8] Michael Bristow. 2008. China's internet 'spin doctors'. *BBC News* (December 2008). http://news.bbc.co.uk/2/hi/asia-pacific/7783640.stm

[9] Morgan Burcham, Mahran Al-Zyoud, Jeffrey C Carver, Mohammed Alsaleh, Hongying Du, Fida Gilani, Jun Jiang, Akond Rahman, Ozgur Kafali, Ehab Al-Shaer, and Laurie Williams. 2017. Characterizing Scientific Reporting in Security Literature: An analysis of ACM CCS and IEEE S&P Papers. In *Proceedings of the Hot Topics in Science of Security: Symposium and Bootcamp*. ACM, 13–23.

[10] Eoghan Casey. 2011. *Digital Evidence and Computer Crime*. Academic Press.

[11] Amy Chang. 2014. *Warring State: China's Cybersecurity Strategy*. Technical Report. Center for a New American Security.

[12] Crispin Cowan, Seth Arnold, Steve Beattie, Chris Wright, and John Viega. 2003. Defcon Capture the Flag: Defending vulnerable code from intense attack. In *Proceedings of DARPA Information Survivability Conference and Exposition 2003*, Vol. 1. IEEE, 120–129.

[13] Fernando de la Cuadra. 2007. The Geneology of Malware. *Network Security* (2007).

[14] Defense Acquisition University. [n. d.]. Research, Development, Test, and Evaluation Budget Activities. https://dap.dau.mil/glossary/pages/2575.aspx Last checked August 21, 2017.. ([n. d.]).

[15] Department of Defense. 2017. Dictionary of Military and Associated Terms. (June 2017). http://www.dtic.mil/doctrine/dod_dictionary/ Last accessed July 11, 2017.

[16] Johanna Drucker. 2009. *Speclab: Digital Aesthetics and Projects in Speculative Computing*. The University of Chicago Press.

[17] Manuel Egele, Theodoor Scholte, Engin Kirda, and Christopher Kruegel. 2012. A survey on automated dynamic malware-analysis techniques and tools. *ACM Computing Surveys (CSUR)* 44, 2 (2012).

[18] Assafa Endeshaw. 2004. Internet regulation in China: the never-ending cat and mouse game. *Information & Communications Technology Law* 13, 1 (2004), 41–57.

[19] David Evans. 2008. Worshop Report: NSF/IARPA/NSA Workshop on the Science of Security. (2008).

[20] Daniel Fendel and Diane Resek. 1990. *Foundations of Higher Mathematics: Exploration and Proof*. Addison Wesley.

[21] James Fieser and Bradley Dowden (editors). [n. d.]. Internet Encyclopedia of Philosophy. http://www.iep.utm.edu/ Last checked August 18, 2017.. ([n. d.]).

[22] Ulrik Franke. 2015. *War by non-military means: Understanding Russian information warfare*. Technical Report FOI-R–4065–SE. Swedish Defence Research Agency (FOI).

[23] Michael Franz. 2015. From Fine Grained Code Diversity to JIT-ROP to Execute-Only Memory: The Cat and Mouse Game Between Attackers and Defenders Continues. In *Proceedings of the Second ACM Workshop on Moving Target Defense*. ACM, 1–1.

[24] Peter Galison. 1994. The Ontology of the Enemy: Norbert Wiener and the Cybernetic Vision. *Critical Inquiry* 21, 1 (1994), 228–266.

[25] Peter Galison. 2012. Augustinian and Manichaean Science. (2012). https://cps-vo.org/file/6418/download/16361 Keynote presentation to the 2012 Science of Security (SoS) Community Meeting, available at https://cps-vo.org/file/6418/download/16361.

[26] Mark A Gallagher and Michael Horta. 2013. Cyber joint munitions effectiveness manual (JMEM). *M&S Journal* 8, 2 (2013).

[27] Donald L. Giadrosich. 1995. *Operations research analysis in test and evaluation*. American Institute of Aeronautics & Astronautics.

[28] Keir Giles and William Hagestad II. 2013. Divided by a Common Language: Cyber Definitions in Chinese, Russian and English. In *5th International Conference on Cyber Conflict*. NATO CCD COE Publications, Tallinn, Estonia.

[29] Donald I Good. [n. d.]. The Foundations of Computer Security: We Need Some. Circulated in 1986, available at http://www.ieee-security.org/CSFWweb/goodessay.html Last checked August 25, 2017.. ([n. d.]).

[30] C.S. Gray. 2013. *Making Strategic Sense of Cyber Power: Why the Sky is Not Falling*. Technical Report. Army War College, Strategic Studies Institute.

[31] Thomas X Hammes. 2009. Information Warfare. In *Ideas as weapons: influence and perception in modern warfare*, G J David and T R McKeldin (Eds.). Potomac Books, Inc., 27–34.

[32] Michael I. Handel. 2001. *Masters of War: Classical Strategic Thought* (3rd ed.). Routledge.

[33] Eric Hatleback. 2017. The Protoscience of Cybersecurity. *Journal of Defense Modeling and Simulation* (2017).

[34] Eric Hatleback and Jonathan M Spring. 2014. Exploring a mechanistic approach to experimentation in computing. *Philosophy & Technology* 27, 3 (2014), 441–459.

[35] Cormac Herley and P C van Oorschot. 2017. SoK: Science, Security, and the Elusive Goal of Security as a Scientific Pursuit. In *Proceedings of the IEEE Symposium on Security and Privacy*.

[36] Adele E Howe, Indrajit Ray, Mike Roberts, Malgorzata Urbanska, and Zinta Byrne. 2012. The psychology of security for the home computer user. In *IEEE Symposium on Security and Privacy*. 209–223.

[37] Stuart H James and Jon J Nordby (Eds.). 2003. *Forensic Science: an introduction to scientific and investigative techniques*. CRC Press.

[38] JASON. 2010. *Science of Cyber-Security. Report number JSR-10-102*. Technical Report. MITRE corporation.

[39] Magnus Johnsson. 2011. *NATO and the challenge of strategic communication*. NDC Fellowship Monograph. NATO Defence College, Italy.

[40] Patrick Kingsley. 2014. Who is behind ISIS's terrifying online propaganda operation? *The Guardian* (June 23 2014). https://www.theguardian.com/world/2014/jun/23/who-behind-isis-propaganda-operation-iraq

[41] Carl E Landwehr. 2011. Cybersecurity: From engineering to science. *The Next Wave* 19, 2 (2011).

[42] Carl E Landwehr, Alan R Bull, John P McDermott, and William S Choi. 1994. A Taxonomy of Computer Program Security Flaws. *Comput. Surveys* (1994).

[43] Eric V Larson, Richard E Darilek, Daniel Gibran, Brian Nichiporuk, Amy Richardson, Lowell H Schwartz, and Cathryn Q Thurston. 2009. *Foundations of effective influence operations: A framework for enhancing army capabilities*. Technical Report. RAND Arroyo Center, Santa Monica CA.

[44] Ulf Lindqvist and Erland Jonsson. 1997. How to Systematically Classify Computer Security Intrusions. In *Proceedings of 1997 IEEE Symposium on Security and Privacy*.

[45] Tom Longstaff, David Balenson, and Mark Matties. 2010. Barriers to science in security. In *Proceedings of the 26th Annual Computer Security Applications Conference*. ACM, 127–129.

[46] Dave MacEslin. 2006. Methodology for Determining EW JMEM. *IO Sphere: The Professional Journal of Joint Information Operations* (Spring 2006).

[47] Peter Machamer, Lindley Darden, and Carl F Craver. 2000. Thinking about mechanisms. *Philosophy of science* 67, 1 (2000), 1–25.

[48] Mandiant Corporation. 2013. *APT1: exposing one of China's cyber espionage units*. Mandiant Corporation.

[49] Roy Maxion. 2011. Making experiments dependable. *Dependable and Historic Computing* (2011), 344–357.

[50] Roy A Maxion, Thomas A Longstaff, and John McHugh. 2010. Why is there no science in cyber science?: a panel discussion at NSPW 2010. In *Proceedings of the 2010 workshop on New security paradigms*. ACM, 1–6.

[51] John McLean. 2014. The Science of Computer Security: Perspectives and Prospects. (Keynote Presentation). In *The 2014 Symposium and Bootcamp on the Science of Security (HotSoS)*.

[52] Thanh Hong Nguyen, Debarun Kar, Matthew Brown, Arunesh Sinha, Albert Xin Jiang, and Milind Tambe. 2016. Towards a science of security games. In *Mathematical Sciences with Multidisciplinary Applications*. Springer, 347–381.

[53] Rishab Nithyanand, Sheharbano Khattak, Mobin Javed, Narseo Vallina-Rodriguez, Marjan Falahrastegar, Emiliano De Cristofaro Julia E Powles and, Hamed Haddadi, and Steven J Murdoch. 2016. Ad-blocking and counter blocking: A slice of the arms race. In *6th USENIX Workshop on Free and Open Communications on the Internet (FOCI 16)*. USENIX Association.

[54] Robert Nola and Howard Sankey. 2007. *Theories of Scientific Method*. Trowbridge.

[55] ODNI. 2017. *Assessing Russian Activities and Intentions in Recent US Elections*. Intelligence Community Assessment ICA 2017-01D. Office of the Director of National Intelligence. https://www.dni.gov/files/documents/ICA_2017_01.pdf declassified version.

[56] Peter Paret, Gordon A Craig, and Felix Gilbert (Eds.). 1986. *Makers of modern strategy from Machiavelli to the Nuclear Age*. Princeton University Press.

[57] Barbara L Peckarsky, Peter A Abrams, Daniel I Bolnick, Lawrence M Dill, Jonathan H Grabowski, Barney Luttbeg, John L Orrock, Scott D Peacor, Evan L Preisser, Oswald J Schmitz, et al. 2008. Revisiting the classics: considering nonconsumptive effects in textbook examples of predator–prey interactions. *Ecology* 89, 9 (2008), 2416–2425.

[58] John S Przemieniecki. 2000. *Mathematical methods in defense analyses*. American Institute of Aeronautics & Astronautics.

[59] Stephen Ramsay. 2011. *Reading machines: Toward an algorithmic criticism*. University of Illinois Press.

[60] Gregory Rattray and Jason Healey. 2010. Categorizing and understanding offensive cyber capabilities and their use. In *Proceedings of a Workshop on Deterring CyberAttacks: Informing Strategies and Developing Options for US Policy*.

[61] Thomas Rid. 2012. Cyber war will not take place. *Journal of Strategic Studies* 35, 1 (2012), 5–32.

[62] Ronald L. Rivest. 1990. Cryptography. In *Handbook of Theoretical Computer Science*, J. Van Leeuwen (Ed.). Vol. 1. Elsevier, Chapter 13, 717–755.

[63] Antonio Roque, Kevin B. Bush, and Christopher Degni. 2016. Security is about Control: Insights from Cybernetics. In *Proceedings of Symposium and Bootcamp*

*on the Science of Security: HotSoS '16*.

[64] Christian Rossow, Christian J Dietrich, Chris Grier, Christian Kreibich, Vern Paxson, Norbert Pohlmann, Herbert Bos, and Maarten Van Steen. 2012. Prudent practices for designing malware experiments: Status quo and outlook. In *2012 IEEE Symposium on Security and Privacy*. IEEE.

[65] Viola Schiaffonati and Mario Verdicchio. 2014. Computing and Experiments: A Methodological View on the Debate on the Scientific Nature of Computing. *Philosophy & Technology* 27 (2014).

[66] Andreas Schmidt. 2012. At the boundaries of peer production: the organization of Internet security production in the cases of Estonia 2007 and Conficker. *Telecommunications Policy* (2012).

[67] Fred B Schneider. 2012. Blueprint for a Science of Cybersecurity. *The Next Wave* 19, 2 (2012).

[68] Krister Segerberg, John-Jules Meyer, and Marcus Kracht. [n. d.]. The Logic of Action. https://plato.stanford.edu/entries/logic-action/ Last checked August 29, 2017.. ([n. d.]).

[69] Paulo Shakarian, Jana Shakarian, and Andrew Ruef. 2013. *Introduction to cyber-warfare: a multidisciplinary approach*. Elsevier.

[70] Jonathan M Spring and Eric Hatleback. 2017. Thinking about intrusion kill chains as mechanisms. *Journal of Cybersecurity* (2017), 1–13.

[71] Matthias Steup. [n. d.]. Epistemology. https://plato.stanford.edu/entries/epistemology Last checked August 18, 2017.. ([n. d.]).

[72] Laszlo Szekeres, Mathias Payer, Tao Wei, and Dawn Song. 2013. SoK: Eternal war in memory. In *Security and Privacy (SP), 2013 IEEE Symposium on*. IEEE, 48–62.

[73] Hamdy A Taha. 2007. *Operations Research: An Introduction* (8th ed.). Pearson Prentice Hall.

[74] US-CERT. [n. d.]. US-CERT Home and Business. https://www.us-cert.gov/home-and-business last checked August 16 2017. ([n. d.]).

[75] Vilhelm Verendel. 2009. Quantified security is a weak hypothesis: a critical survey of results and assumptions. In *Proceedings of the 2009 New security paradigms workshop*. ACM, 37–50.

[76] Verizon. 2014. *2014 Data Breach Investigation Report*. Technical Report. Verizon.

[77] Glenn Voelz. 2014. Is Military Science "Scientific"? *Joint Force Quarterly* (2014).

[78] Carl von Clausewitz. 1989. *On War*. Princeton University Press. Edited and translated by Michael E. Howard and Peter Paret.

[79] John Von Neumann. 1959. On the theory of games of strategy. *Contributions to the Theory of Games* 4 (1959), 13–42.

[80] Willis H Ware. 1970. *Security Controls for Computer Systems (U): Report of Defense Science Board Task Force on Computer Security*. Technical Report. RAND Corp.

[81] Norbert Wiener. 1989, (First Published 1950 by Houghton Mifflin). *The Human Use of Human Beings*. Free Association Books.

[82] Wikipedia. [n. d.]. Turtles all the way down. https://en.wikipedia.org/wiki/Turtles_all_the_way_down Last checked August 16, 2017. ([n. d.]).